\documentclass[aoas,MSNbibl,nameyear,dvips]{arximspdf}
\usepackage{algorithm}
\usepackage{algorithmic}
\usepackage{dcolumn}
\usepackage{graphicx}

\doi{10.1214/11-AOAS504}
\volume{6}
\issue{1}
\pubyear{2012}
\firstpage{253}
\lastpage{283}

\makeatletter

\newcolumntype{d}[1]{D{.}{.}{#1}}

\newcommand{\reals}{{\mathbf R}}

\newcommand{\by}{\mathbf{y}}
\newcommand{\bhy}{\hat{\mathbf{y}}}
\newcommand{\bx}{{\mathbf x}}
\newcommand{\hy}{\hat{y}}
\newcommand{\mC}{\mathcal{C}}
\newcommand{\tr}{\mathsf{T}}
\newcommand{\mE}{\mathbb{E}}

\newcommand{\ei}{\mathrm{ERR}_{\mathrm{in}}}
\newcommand{\byn}{\by^{\mathrm{new}}}
\newcommand{\df}{\mathit{df}}

\newtheorem{theorem}{Theorem}
\newtheorem{proposition}[theorem]{Proposition}
\newtheorem{corollary}[theorem]{Corollary}
\newproclaim{remark}[theorem]{Remark}

\makeatother

\begin{document}
\begin{frontmatter}

\title{Efficient regularized isotonic regression with application to gene--gene interaction search}
\runtitle{Efficient regularized isotonic regression}

\begin{aug}
\author[A]{\fnms{Ronny} \snm{Luss}\corref{}\thanksref{t1}\ead[label=e1]{ronnyluss@gmail.com}},
\author[A]{\fnms{Saharon} \snm{Rosset}\thanksref{t1}\ead[label=e2]{saharon@post.tau.ac.il}}
\and
\author[B]{\fnms{Moni} \snm{Shahar}\ead[label=e3]{moni@eng.tau.ac.il}}
\runauthor{R. Luss, S. Rosset and M. Shahar}
\affiliation{Tel Aviv University}
\address[A]{R. Luss\\
S. Rosset\\
School of Mathematical Sciences\\
Tel Aviv University\\
Ramat Aviv, Tel Aviv, 69978\\
Israel\\
\printead{e1}\\
\phantom{E-mail: }\printead*{e2}}
\address[B]{M. Shahar\\
Faculty of Engineering\\
Tel Aviv University\\
Ramat Aviv, Tel Aviv, 69978\\
Israel\\
\printead{e3}} 
\end{aug}

\thankstext{t1}{Supported in part by
Israeli Science Foundation Grant 1227/09 and an IBM Open Collaborative
Research grant.}

\received{\smonth{2} \syear{2011}}
\revised{\smonth{8} \syear{2011}}

%
\begin{abstract}
Isotonic regression is a nonparametric approach for fitting monotonic
models to data that has been widely studied from both theoretical and
practical perspectives. However, this approach encounters
computational and statistical overfitting issues in higher dimensions.
To address both concerns, we present an algorithm, which we term
Isotonic Recursive Partitioning (IRP), for isotonic regression based on
recursively partitioning the covariate space through solution of
progressively smaller ``best cut'' subproblems. This creates a
regularized sequence of isotonic models of increasing model complexity
that converges to the global isotonic regression solution. The models
along the sequence are often more accurate than the unregularized
isotonic regression model because of the complexity control they offer.
We quantify this complexity control through estimation of degrees of
freedom along the path. Success of the regularized models in prediction
and IRPs favorable computational properties are demonstrated through a
series of simulated and real data experiments. We discuss application
of IRP to the problem of searching for gene--gene interactions and
epistasis, and demonstrate it on data from genome-wide association
studies of three common diseases.
\end{abstract}

%

\end{frontmatter}

\section{Introduction} \label{sintroduction}
In predictive modeling, we are given a set of $n$ data observations
$(x_1, y_1),\ldots,(x_n,y_n)$, where $x\in\mathcal X$ (usually
${\mathcal
X}={\mathbb R}^d$) is a vector of covariates or independent
variables, $y \in\mathbb R$ is the response, and we wish to fit a
model $\hat{f}\dvtx\mathcal X \rightarrow\mathbb R$ to describe the
dependence of $y$ on $x$. 
The most common approach traditionally used in statistics to
accomplish this task is to seek a model that minimizes in-sample squared
error ($l_2$) loss, that is,
\[
\hat{f} =\mathop{\arg\min}_{f \in\mathcal F} \sum_{i=1}^n
\bigl(y_i-f(x_i)\bigr)^2,
\]
where $\mathcal F$ is a class of candidate models. The use of squared
error loss can be interpreted as maximum likelihood fitting under an
additive Gaussian noise assumption, or, more generally, as trying to
estimate $\mathbb E(y|x)$ [\citet{Gnei2011}]. Other loss
functions can be used to
represent other estimation goals.

Isotonic regression is a nonparametric modeling approach which
restricts the fitted model to being monotone in all independent
variables [\citet{Barl1972}]. Define $\mathcal G$ to be the
family of
isotonic functions, that is, $g\in\mathcal G$ satisfies
\[
x_1 \preceq x_2 \quad\Rightarrow\quad g(x_1) \leq g(x_2),
\]
where the partial order $\preceq$ here will usually be the standard
Euclidean one, that is, $x_1 \preceq x_2$ if and only if $x_{1j} \leq
x_{2j}$ coordinate-wise. Given these definitions, standard $l_2$
isotonic regression solves
%
\begin{equation}\label{isomodel}
\min_{g\in\mathcal G}{\sum_{i=1}^n \bigl(y_i - g(x_i)\bigr)^2}.
\end{equation}
We denote by $\hat{f}$ the optimal solution to (\ref{isomodel}). As
many authors have noted,~$\hat{f}$ comprises a partitioning of the
space $\mathcal X$ into regions with no ``holes'' that satisfy isotonicity
properties defined below, with a constant fitted to $\hat{f}$ in
every region.

In terms of model form, isotonic regression is clearly very attractive
in situations where monotonicity is a reasonable assumption, but other
common assumptions like linearity or additivity are not. Indeed, this
formulation has found useful applications in biology
[\citet{Oboz2008}], medicine [\citet {Sche1997}], statistics
[\citet{Barl1972}] and psychology [\citet {Kruskal64}], among
others. In recent years, an exciting new application area has emerged
for this approach in genetics: modeling genetic interactions in
heritability. Many papers have noted the apparent insufficiency of
standard additive modeling approaches in describing the combined
effects of genetic factors (e.g., mutations) on phenotypes like height
and disease [\citet{Goldstein2009}, \citet{Eichler2010}].
Some findings in mice have pointed to subadditive interactions
[\citet{Shao2008}], while others suggest requiring super-additive
assumptions in order to explain heritability
[\citet{Goldstein2009}]. It is generally accepted, however, that
while the effect of one genetic factor on a phenotype can be modulated,
enhanced or even eliminated by other genetic factors, it is not
expected to reverse direction [\citet{Mani2007},
\citet{Roth2009}]. In other words, the isotonicity assumption with
respect to genetic effects is widely accepted, but the form of
epistasis (genetic interaction) between factors is not clear and may
vary between phenotypes. Other properties of this application domain
also favor the use of isotonic regression as we discuss below. It
should be noted, however, that most discussions of epistasis have
been
theoretical, with extensive systematic efforts at discovering actual
epistatic combinations in human disease, resulting in very limited\vadjust{\goodbreak}
[\citet{Emily2009}] or no results [\citet{Cordell2009}].
These efforts have utilized simple statistical approaches (chi-square
tests, logistic regression) to search for low-dimensional interactions,
mostly of two mutations at a time. The question of whether their lack
of findings is due to limitations of methodology and concentration on
low dimension, or lack of real epistatic signal, is a key one, and it
can be addressed by isotonic modeling.

Two major concerns arise when considering the practical use of
isotonic regression in \textit{modern} situations as the number of
observations $n$, the data dimen\-sionality~$d$, and the number of
isotonicity constraints $m=|\{(i,j)\dvtx\break x_i\preceq x_j\}|$ implied by
(\ref{isomodel}) all grow large: statistical overfitting and
computational difficulty. The notations $n$, $d$ and $m$ will
refer to these quantities throughout the paper.

The first concern is statistical difficulty and overfitting. Beyond
very low dimensions, the isotonicity constraints on the family
$\mathcal
G$ can become inefficient in controlling model complexity and the
isotonic regression solutions can be severely overfitted [e.g.,
see \citet{Bacc1989} and \citet{Sche1997}]. At the extreme,
there may be no isotonicity constraints because no two observations
obey the coordinate-wise requirement for the $\preceq$ ordering. The
isotonic\vspace*{1pt} solution in this case simply assigns $\hat{f}(x_i) = y_i$,
providing a perfect interpolation of the training data. As
demonstrated in the literature [\citet{Sche1997}] and below, the
overfitting concern is clearly well-founded when considering the
optimal isotonic regression model implied by~(\ref{isomodel}), even
in nonextreme cases with a large number of constraints. In this
case, regularization, that is, fitting isotonic models that are
constrained to a restricted subset of $\mathcal G$, could offer an
approach that maintains isotonicity while controlling variance,
leading to improved accuracy.

A second concern is computational difficulty. The discussion of
isotonic regression originally focused on the case $x \in\mathbb R$,
where $\preceq$ denoted a complete order [\citet {Kruskal64}]. For
this case, the well-known pooled adjacent violators algorithm (PAVA)
efficiently solves (\ref{isomodel}) in computational complexity~$O(n)$.
Low complexities can also be found when the isotonic constraints take a
special structure such as a tree [$O(n\log{n})$ in
\citet{Pard1999}]. Various algorithms have been developed for the
partially ordered case, including the classical approach of
\citet{Dyks1982} for data on a grid, generalizations of PAVA
[\citet{Lee1983}, \citet{Bloc1994}] and active set methods
[\citet{Leeu2009}]. These approaches offer no polynomial
complexity guarantees and are impractical when data sizes exceed a few
thousand observations (in some cases much less). Interior point methods
offer complexity guarantees of $O(\max{(m,n)}^3)$
[\citet{Mont1989b}], however, they are impractical for large data
sizes due to excessive memory requirements.
A much more computationally attractive approach can be found in the
optimization and operations research literature. The basic idea of
this approach is to repeatedly and ``optimally'' split the covariate
space $\mathcal X$ into regions of decreasing size by solving a sequence
of specially structured \textit{best cut} problems for which efficient
algorithms exist. At most $n$ partitions are needed, leading to a
computational complexity bounded by~$O(n^4)$, and in some cases even
less. From a practical performance perspective, this algorithm can
obtain an exact solution of (\ref{isomodel}) for data sets with tens
of thousands of observations in minutes. The first appearance of
this approach, to our knowledge, is in the work of \citet{Maxw1985}
[and similarly \citet{Round1986}], who presented a model for finding
reorder intervals in a production-distribution system. At the
center of their problem was a regression subject to isotonicity
constraints, but with a different loss function than that in problem
(\ref{isomodel}), to which they provided an algorithm based on
partitioning. Applicability of their algorithm with minimal changes
to problem (\ref{isomodel}) was more recently noticed by several
authors [e.g., \citet{Spou2003}], who used it to state a similar
efficient partitioning scheme for isotonic regression. A similar,
highly efficient, algorithm by \citet{Hoch2003} also solves
problem (\ref{isomodel}) under the additional constraint that fits
take integer values. This partitioning approach
does not appear to be well known in the statistics community, and,
indeed, we have independently developed it before discovering it is
already known.

The literature cited above invariably refers to this iterative
splitting algorithm merely as an approach for efficiently arriving
at the optimal solution of (\ref{isomodel}). However, as noted
before, this solution can be highly overfitted, especially as the
dimension $d$ increases. Our main interest lies in analyzing the
iterative approach as a means toward resolving the overfitting
problem, as well as the computational issue. We propose to view
this iterative algorithm as a \textit{recursive partitioning} approach
that generates isotonic models of increasing model complexity,
ultimately leading to the solution of (\ref{isomodel}); the
algorithm is termed Isotonic Recursive Partitioning (IRP). We prove
that the models generated by the IRP iterations are indeed isotonic
(Theorem \ref{thisotonicsolutions}) and consider them as a
\textit{regularization path} of increasingly complex isotonic regression
models. Models along the path are less complex, and hence likely to be
less overfit and offer better predictive performance than the
overall solution to (\ref{isomodel}), while still maintaining
isotonicity. This is confirmed by our analysis of the equivalent
degrees of freedom along the IRP path, as well as experiments with
simulated and real data.

We observe that for very low dimension (typically $d\leq2$) the
nonregularized solution of (\ref{isomodel}) performs well. As the
dimension increases, regularization becomes necessary, and
intermediate models on the IRP path perform better than the
nonregularized solution. However, eventually overfitting plagues
IRP from its first iteration, and the isotonic models fail to
perform better than simple linear regression in out-of-sample
prediction, even when the linear model is inappropriate. In our
simulations, this occurs around dimensions~6--8 even for relatively
large data sets.\vadjust{\goodbreak}

\begin{figure}

\includegraphics{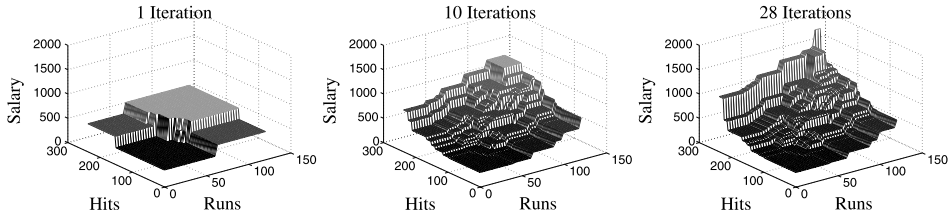}

\caption{Illustration of IRP on Baseball data. Salary is modeled by
number of runs batted in and hits. Models after iterations 1 and 10
of IRP and the final model are shown.} \label{figbaseball}
\end{figure}

Progress of IRP is illustrated in Figure \ref{figbaseball}, where we
show an example of applying IRP to the well-known Baseball data set
[\citet{He1998}] describing the dependence of salary on a
collection of player properties. We limit the model to only two
covariates to facilitate visualization, and we choose to use the number
of runs batted in and hits since they seemed a~priori most likely to
comply with the isotonicity assumptions. The increasing model
complexity can be seen, moving from iteration 1 (a single split)
through 10 iterations of IRP, to the final isotonic model optimally
solving~(\ref{isomodel}), comprising a splitting of the covariate space
into 29 regions, each of which is fitted with a constant. Note that the
single split from iteration 1 creates two flat surfaces (the highest
and lowest in the figure), and that the \textit{in-between} surfaces
are interpolations in regions with no data points based on the
two-surface model. The figures for models after 10 and~28 iterations
are quite complex for the same reason---interpolations in the
continuous covariate space.

%

An obvious analogy of IRP can be made to well-known recursive
partitioning approaches for regression such as CART [\citet{CART}],
where the iterative splitting of the covariate space generates a
sequence of models (trees) of increasing model complexity, from
which the ``best'' tree is chosen via cross-validation [e.g.,
using the 1-SE rule \citet{CART}]. As with CART and other similar
approaches, IRP performs a greedy search and finds a ``local''
optimum in every iteration. However, unlike CART, which has no
guarantees on the overall model it generates, IRP is proven to
terminate in the global solution of the isotonic regression
problem (\ref{isomodel}). Another difference is that IRP splits are
not made along one axis at a time, but rather each split is a
nonparametric division of one region in $\mathcal X$ into two
subregions.

Importantly, while our presentation has so far concentrated on isotonic
regression with an $l_2$ loss function, an interesting extension of
the partitioning scheme can be used to solve large-scale isotonic
regressions under other loss functions that are of great interest in
statistics, for example, logistic and poisson log-likelihoods.
Specifically, a well-known result of Barlow and Brunk [(\citeyear{Barl1972}),
Theorem 3.1],
implies that, for a large class of loss functions (formally
defined\vadjust{\goodbreak}
in our Discussion section), the solution of isotonic mode\-ling can be found
by a simple transformation of the solution to isotonic regression
with an $l_2$ loss function. This will play an important role in our
use of isotonic regression on modeling gene--gene interactions in
epistasis. Indeed, there we will maximize logistic log-likelihood
subject to isotonicity constraints.

The remainder of this paper is organized as follows. We first
present and analyze the IRP algorithm in Section \ref{salg}. We
detail the best cut problem solved for splitting at each iteration,
and prove that this algorithm is a~\textit{no-regret} algorithm, in
the sense that it only partitions the data and never merges back
previously made partitions and converges to the global solution of
(\ref{isomodel}) (Theorem \ref{thnoregretcut}). Furthermore, we
prove that the intermediate partitions generated along the IRP path
are also isotonic, in the sense that fitting each region to the average
gives a model that is in the class $\mathcal G$ of isotonic
functions in $\mathbb R^d$ (Theorem \ref{thisotonicsolutions}).
Section \ref{scomplexity} briefly reviews the theoretical
computational guarantees of IRP as reflected in the literature, and
develops a simple and realistic case where the overall computation
is $O(n^3)$. Section \ref{sdegreesfreedom} discusses the
statistical model complexity of models generated along the
regularization path. \citet{Meyer2000} have shown that the number of
partitions in the solution of (\ref{isomodel}) is an unbiased
estimator of the (equivalent) degrees of freedom [as defined by
\citet{Efron1986}]. Since IRP adds one to the number of partitions
at each iteration, the number of iterations may be used as a
parametrization of this sequence. However, we argue that the number
of regions is not a good estimate of degrees of freedom because IRP
performs much more fitting in its initial iterations compared to
later stages, and demonstrate this effect empirically through
simulation. We also show that when the covariates are ternary, that is,
covariate $x_{ij}\in\{0,1,2\}$ (as is natural in our motivating
genetic example when dealing with ternary genotype data), the
overall number of degrees of freedom and model complexity increase
more slowly with dimension, compared to general continuous
covariates, resulting in much less overall fitting for each
dimension. Section~\ref{sperformance} examines IRPs statistical
and computational performance on simulated and real data,
specifically pointing out the effect of regularization and increased
dimensionality on predictive performance. We apply IRP to
simulations with ternary covariates and sub- and super-additive
interactions motivated by the genetic application and demonstrate
its favorable performance. Section \ref{sepistasis} applies IRP to
real-life data sets from the Wellcome Trust Case Control Consortium
(WTCCC) and discusses the results  [\citet{WTCCC2007}].
Section \ref{sdiscussion} concludes with extensions and connections
to previous literature. A Matlab-based software package implementing
the IRP algorithm is available at
\url{http://www.tau.ac.il/\textasciitilde saharon/files/IRPv1.zip}.

We next define terminology to be used throughout the paper.

\subsection{Definitions}
Let $V=\{x_1,\ldots,x_n\}$ be the covariate vectors for $n$ training
points where
$x_i\in\mathbb R^d$ and denote $y_i\in\mathbb R$ as the $i$th
observed\vadjust{\goodbreak} response. We will refer to a general subset of
points $A\subseteq V$ with no holes (i.e., $x\preceq y\preceq z$ and
$x,z\in A\Rightarrow y\in A$) as a \textit{group}. Denote by $|A|$ the
cardinality of group $A$. The \textit{weight}
of group $A$ is defined as $\overline{y}_A=\frac{1}{|A|}\sum
_{i\dvtx x_i\in A}{y_i}$. For two groups $A$ and $B$, we
denote $A\preceq B$ if there exists $x\in A, y\in B$ such that
$x\preceq y$
and there does not exist $x\in A, y\in B$ such that $y\prec x$ (i.e.,
there is at
least one comparable pair of points that satisfy the direction of
isotonicity). A set of groups $\mathcal{V}$ is called isotonic if
$A\preceq B\Rightarrow\overline{y}_A\leq\overline{y}_B$ for all
$A,B\in\mathcal{V}$. The groups within this set $\mathcal{V}$ are
referred to as isotonic regions. A~subset~$\mathcal{L}$
($\mathcal{U}$) of $A$ is a \textit{lower set} (\textit{upper set}) of
$A$ if $x\in A,y\in\mathcal{L},x\prec y\Rightarrow x\in\mathcal{L}$
($x\in\mathcal{U},y\in A,x\prec y\Rightarrow y\in\mathcal{U}$).

A group $B\subseteq A$ is defined as a block of group $A$ if
$\overline{y}_{\mathcal{U}\cap B}\leq\overline{y}_B$ for each upper
set $\mathcal{U}$ of $A$ such that $\mathcal{U}\cap B \ne\{\cdot\}$ (or,
equivalently, if $\overline{y}_{\mathcal{L}\cap B}\geq
\overline{y}_B$ for each lower set $\mathcal{L}$ of $A$ such that
$\mathcal{L}\cap B \ne\{\cdot\}$). A set of blocks $\mathcal{S}=\{
B_1,\ldots,B_k\}$ is
called a \textit{block class} of $V$ if $B_i\cap B_j=\{\cdot\}$ and
$B_1\cup\cdots\cup B_k=V$. $\mathcal{S}$ is an
\textit{isotonic block class} if for all $B_i,B_j\in\mathcal{S}$,
$B_i\preceq B_j\Rightarrow\overline{y}_{B_i}\leq\overline
{y}_{B_j}$. A group $X$ \textit{majorizes} (\textit{minorizes}) another
group $Y$ if
$X\succeq Y$ ($X\preceq Y$). A group $X$ is a~\textit{majorant}
(\textit{minorant}) of $X\cup A$ where $A=\bigcup_{i=1}^k{A_i}$ if
$X\not\prec A_i$ ($X\not\succ A_i$) for all $i=1,\ldots, k$.

We denote the optimal solution for maximizing a general function $f(z)$
in the variable $z$ by $z^*$, that is, $z^*=\arg\max{f(z)}$.\vspace*{-3pt}

\section{IRP and a regularization path for isotonic regression} \label{salg}
We describe here the partitioning algorithm used to solve the isotonic
regression problem (\ref{isomodel}). Section \ref{ssstructure} first
reformulates the isotonic regression problem and describes the
structure of the optimal solution. Section \ref{sssplittingalgo}
motivates and details the IRP algorithm and, in particular, the main
partitioning step. Each group created by the partitioning scheme is
proven to be the union of blocks in the optimal solution, that is, all
partitions have the no-regret property. An important aspect of the
algorithm is the regularization path generated as a byproduct as each
partition creates a new feasible solution. Section \ref{sspathconv}
goes on to prove convergence of IRP to the global optimal solution of
(\ref{isomodel}), and most importantly, that each solution along the
regularization path is isotonic.\vspace*{-3pt}

\subsection{Structure of the isotonic solution} \label{ssstructure}
Isotonic regression seeks a monotonic function that fits a given
training data set $\{(x_i,y_i)\}_{i=1}^n$ and satisfies a~set of
\textit{isotonicity constraints} which we index by the set $\mathcal
{I}=\{
(i,j)\dvtx x_i \preceq x_j\}$. We
will usually assume that $x_i \in\mathbb R^d$ and that $\preceq$ is
the standard partial order in $\mathbb R^d$ based on coordinate-wise
inequalities. A reformulation of (\ref{isomodel}) is
%
\begin{equation}\label{eqIR}
\min\Biggl\{ \sum_{i=1}^n{(\hat{y}_i-y_i)^2} \dvtx\hat
{y}_i\leq\hat{y}_j \mbox{ for all } (i,j)\in\mathcal
{I}\Biggr\}.
\end{equation}
Problem (\ref{eqIR}) is a quadratic program with linear constraints.
Any solution satisfying the constraints\vadjust{\goodbreak} given by $\mathcal{I}$ is
referred to as an isotonic, or feasible, solution. The structure of the
optimal solution to (\ref{eqIR}) is well known. Observations are
divided into $k$ groups where
the fits in each group take the group mean observation value. This can
be seen through the following Karush--Kuhn--Tucker (KKT), that is,
optimality, conditions [\citet{Boyd2004}] to (\ref{eqIR}):
\begin{longlist}[(a)]
\item[(a)]$ \hat{y}_i=y_i-\frac{1}{2}(\sum_{j\dvtx(i,j)\in
\mathcal{I}}{\lambda_{ij}}-\sum_{j\dvtx(j,i)\in\mathcal{I}}{\lambda_{ji}})$,
\item[(b)]$\hat{y}_i\leq\hat{y}_j$ for all $(i,j)\in\mathcal{I}$,
\item[(c)]$\lambda_{ij}\geq0$ for all $(i,j)\in\mathcal{I}$,
\item[(d)]$\lambda_{ij}(\hat{y}_i-\hat{y}_j)=0$ for all $(i,j)\in
\mathcal{I}$,
\end{longlist}
where $\lambda_{ij}$ is the dual variable corresponding to the
isotonicity constraint $\hat{y}_i\leq\hat{y}_j$. This set of
conditions exposes the nature of the optimal solution. Condition (d)
implies that $\lambda_{ij} > 0 \Rightarrow\hy_i
= \hy_j$, meaning $\lambda_{ij}$ can be nonzero only within blocks in
the isotonic solution which have the same fitted value. For
observations in different blocks, $\lambda_{ij}=0$. Furthermore,
the fit within each block is trivially seen to be the average of the
observations in the block, as the average minimizes the block's
squared loss. A block is thus also referred to as an \textit{optimal
group} with respect to an isotonic regression problem. Condition (b)
implies isotonicity of the blocks, and thus, we get the familiar
characterization of the isotonic regression problem as one of finding a
division into an isotonic block class.

\subsection{The partitioning algorithm} \label{sssplittingalgo}
Suppose a current group $V$ is optimal (i.e., $V$ is a block) and,
thus, the optimal fits at points in $V$, denoted $\hat{y}_i^*$,
satisfy $\hat{y}_i^*=\overline{y}_V$ for all $i\in V$, which leads to
the condition $\sum_{i\in V}{(y_i-\overline{y}_V)}=0$. Then finding
two groups $A$ and $B$ within $V$ such that $\sum_{i\in
B}{(y_i-\overline{y}_V)}-\sum_{i\in A}{(y_i-\overline{y}_A)}>0$
should be infeasible, according to the KKT conditions. The division in
IRP looks for two such groups. Denote by $\mathcal{C}_V=\{
(A,B)\dvtx\break A,B\subseteq V,A\cup B=V,A\cap B=\{\cdot\}$, and there does not
exist $x\in A,y\in B$ such that  $y\preceq x\}$
the set of all feasible (i.e., isotonic) partitions defined by
observations in $V$. We refer to partitioning as making a cut through
the variable space (hence, our optimal partition is made by an
\textit{optimal cut}). The optimal cut is determined by
the partition that solves the problem\looseness=1
%
\begin{eqnarray}\label{eqoptimalcut}
\max_{(A,B)\in\mathcal{C}_V}{g^*(A,B)}&=&\max_{(A,B)\in\mathcal
{C}_V}\biggl\{\sum_{i\in B}
{(y_i-\overline{y}_V)}-\sum_{i\in
A}{(y_i-\overline{y}_V)}\biggr\}\nonumber\\[-8pt]\\[-8pt]
&=& -|A|(\overline{y}_A-\overline{y}_{V})+|B|(\overline
{y}_B-\overline{y}_{V}),\nonumber
\end{eqnarray}\looseness=0
where $A$($B$) is the group on the lower (upper) side of the edges of
the cut. A more statistically intuitive rule might look for the split
that maximizes between-group variance. This partitioning problem solves
%
\begin{equation}\label{eqbetweenvarcut}
\max_{(A,B)\in\mathcal{C}_V}{\tilde{g}(A,B)}=\max_{\{(A,B)\in
\mathcal{C}_V\}}{\{|A|(\overline{y}_A-\overline
{y}_{V})^2+|B|(\overline{y}_B-\overline{y}_{V})^2\}}.
\end{equation}

The next proposition makes a connection between the above two
maximization problems, and draws a clear conclusion on the relationship
between their optimal solutions, namely, that the optimal partitions to
(\ref{eqoptimalcut}) are always more balanced than the optimal
partitions to (\ref{eqbetweenvarcut}).
\begin{proposition} \label{propcartvsirp}
Denote the optimal solutions of the optimal cut problem (\ref
{eqoptimalcut}) and the between-group variance maximization problem
(\ref{eqbetweenvarcut}) by $(A^*,B^*)$ and $(\tilde{A},\tilde{B})$
and their objective functions by $g^*(A,B)$ and $\tilde{g}(A,B)$,
respectively. Then
\[
(A^*,B^*)=\mathop{\arg\max}_{(A,B)\in\mathcal{C}_V}{\{|A||B|\tilde
{g}(A,B)\}}
\]
and
\[
(|A^*|-|B^*|)^2\leq(|\tilde{A}|-|\tilde{B}|)^2.
\]
\end{proposition}

We leave the proof to the \hyperref[app]{Appendix}.

Thus, we can look at the IRP criterion as a modified form of maximizing
between-group variance which encourages more balanced splitting.
However, while solving the partition problem (\ref{eqbetweenvarcut})
is difficult, the IRP partition problem~(\ref{eqoptimalcut}) is
tractable. Indeed, the optimal partition problem (\ref{eqoptimalcut})
can be reduced to solving the linear program
%
\begin{equation}\label{eqoptimalcutlp}
\max{\{c^Tz\dvtx z_i\leq z_j \mbox{ for all } (i,j)\in\mathcal
{I},-1\leq z_i\leq1 \mbox{ for all } i\in V\}},
\end{equation}
where $c_i=y_i-\overline{y}_V$. If the optimal objective value equals
zero, then the group~$V$ must be an optimal block.

\begin{algorithm}[t]
\caption{Isotonic Recursive Partitioning}\label{algIRnoregret}
\begin{algorithmic} [1]
\REQUIRE Observations $(x_1,y_1),\ldots,(x_n,y_n)$ and partial order
$\mathcal{I}$.
\REQUIRE$k=0,\mathcal{A}=\{\{x_1,\ldots,x_n\}\}$, $\mathcal{C}=\{
(0,\{x_1,\ldots,x_n\},\{\cdot\})\}$, $\mathcal{B}=\{\cdot\}$, $M_0=(\mathcal
{A},\overline{y}_\mathcal{A})$.
\WHILE{$\mathcal{A}\ne\{\cdot\}$}
\STATE Let $(\mathit{val},w^-,w^+)\in\mathcal{C}$ be the potential partition
with largest $\mathit{val}$.
\STATE Update $\mathcal{A}=(\mathcal{A}\setminus(w^-\cup w^+))\cup\{
w^-,w^+\}$, $\mathcal{C}=\mathcal{C}\setminus(\mathit{val},w^-,w^+)$.
\STATE$M_k=(\mathcal{A},\overline{y}_{\mathcal{A}})$.
\FORALL{$v\in\{w^-,w^+\}\setminus\{\cdot\}$}
\STATE Set $c_i=y_i-\overline{y}_v$ for all $i$ such that $x_i\in v$
where $\overline{y}_v$ is the mean of the observations in set $v$.
\STATE Solve LP (\ref{eqoptimalcutlp}) with input $c$ and get
$z^*=\arg\max{\mbox{LP}(\ref{eqoptimalcutlp})}$.
\IF{$z_1^*=\cdots=z_n^*$ (group is optimally divided)}
\STATE Update $\mathcal{A}=\mathcal{A}\setminus v$ and $\mathcal
{B}=\mathcal{B}\cup\{v\}$. \ELSE
\STATE Let $v^-=\{x_i\dvtx z^*_i=-1\}, v^+=\{x_i\dvtx z^*_i=+1\}$.
\STATE Update $\mathcal{C}=\mathcal{C}\cup\{(c^Tz^*,v^-,v^+)\}$
\ENDIF
\ENDFOR
\STATE $k=k+1$.
\ENDWHILE
\RETURN$\mathcal{B}$, a partitioning of observations corresponding to
the optimal groups.
\end{algorithmic}
\end{algorithm}

This group-wise partitioning operation is the basis for our IRP
algorithm which is detailed in
Algorithm \ref{algIRnoregret}.\setcounter{footnote}{1}\footnote
{This algorithm appeared in a shorter version of the paper [\citet
{Luss2010b}].} It starts with all observations as one group and
recursively splits each group optimally by solving subproblem~(\ref
{eqoptimalcutlp}). At each iteration, a list
$\mC$ of potential optimal partitions for each group generated thus
far is maintained, and the partition among them with the highest
objective value
is performed. The list $\mC$ is updated with the optimal partitions
generated from
both subgroups. Partitioning ends whenever the solution
to (\ref{eqoptimalcutlp}) is trivial (i.e., no split is found
because the group is a block). We can think of each iteration $k$ of
Algorithm \ref{algIRnoregret} as producing a model~$M_k$ by fitting
the average to each group in its current partition: for a set of groups
$\mathcal{V}=\{V_1,\ldots,V_k\}$, denote $\overline{y}_\mathcal
{V}=\{\overline{y}_{V_1},\ldots,\overline{y}_{V_k}\}$. Then model
$M_k=(\mathcal{V},\overline{y}_\mathcal{V})$ contains the
partitioning $\mathcal{V}$ as well as a fit to each of the
observations, which is the mean observation of the group it belongs to
in the partition.

\subsection{Properties of the partitioning algorithm} \label{sspathconv}
Theorem \ref{thnoregretcut} next states the result which implies that
the IRP partitions are \textit{no-regret}. This will lead to our
convergence result.\vadjust{\goodbreak}
\begin{theorem} \label{thnoregretcut}
Assume group $V$ is the union of blocks from the
optimal solution to problem (\ref{eqIR}). Then a cut made by
solving (\ref{eqoptimalcut}) [using (\ref{eqoptimalcutlp})] at
a~particular iteration does not cut through any block in the global
optimal solution.\looseness=1
\end{theorem}

The fact that IRP is a no-regret algorithm can be shown using a
connection between the work of \citet{Barl1972} and \citet
{Maxw1985} (held to the Discussion in Section \ref{sdiscussion}). We
prove Theorem~\ref{thnoregretcut} directly, but leave it to the
\hyperref[app]{Appendix}, as the theorem is already known to be true [\citet
{Spou2003}]. Remark \ref{remmultipleobservations} in the \hyperref[app]{Appendix}
handles the case for multiple observations. Since Algorithm \ref
{algIRnoregret} starts with $\mathcal{A}=\{\{x_1,\ldots,x_n\}\}$,
which is a set of one group that is the union of all blocks, we can
conclude from this
theorem that IRP never cuts an optimal block when generating
partitions. The following corollary is then a direct consequence of
repeatedly applying Theorem \ref{thnoregretcut} in Algorithm
\ref{algIRnoregret}:
\begin{corollary} \label{corrconvergence}
Algorithm \ref{algIRnoregret} converges to the optimal (isotonic
block class) solution with no regret.
\end{corollary}

Theorem \ref{thisotonicsolutions} next states our main innovative
result that Algorithm \ref{algIRnoregret} provides isotonic solutions
at each iteration. This result implies that the path of solutions
generated by IRP can be regarded as a regularization path for isotonic
regression. Along the path, the model grows in complexity until
optimality. These suboptimal isotonic models often result in better
predictive performance than the optimal solution, which is susceptible
to overfitting as is discussed in Section \ref{sperformance}.
\begin{theorem} \label{thisotonicsolutions}
Model $M_k$ generated after iteration $k$ of Algorithm \ref
{algIRnoregret} is in the class $\mathcal G$ of isotonic models.
\end{theorem}
\begin{pf}
The proof is by induction. The base case, that is, first iteration,
where all points form one group is trivial. The first cut is made
by solving the linear program (\ref{eqoptimalcutlp}) which constrains
the solution to maintain isotonicity.

Assuming that iteration $k$ (and all previous iterations) provides
an isotonic solution, we prove that iteration $k+1$ must also
maintain isotonicity. Figure \ref{figproof2} helps illustrate the
situation described here. Let $G$ be the group split at iteration $k+1$
and denote $A$ ($B$) as the group under (over) the cut. Let
$\mathcal{A}=\{X\dvtx X $ is a group at iteration $k+1$, there exists
$x_i\in X$
such that $(i,j)\in\mathcal{I}$ for some $x_j\in A\}$ (i.e.,
$X\in\mathcal{A}$ border $A$ from below).

%
\begin{figure}

\includegraphics{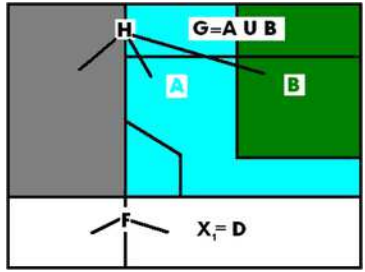}

\caption{Illustration of the proof of Theorem
\protect\ref{thisotonicsolutions} showing the defined sets at
iteration $k+1$. $G$ is the set divided at iteration $k+1$ into $A$
(all blue area) and $B$ (all green area). The group bordering $A$ from
below denoted by $X_1$ (also referred to as $D$ in the proof) is in
violation with $A$. At iteration $k_0$, $G$ is part of the larger group
$H$ and $X_1$ is part of the larger group $F$. At iteration $k_0$,
groups $F$ and $H$ are separated. The proof shows that when~$A$ and~$B$
are split at iteration $k+1$, no group such as $X_1$ where
$w_{X_1}>w_A$ could have existed. In the picture, $X_1$ must have been
separated at an iteration $k_0<k+1$, but the proof, through
contradiction, shows that this cannot occur.} \label{figproof2}
\end{figure}

Consider iteration $k+1$. Denote
$\mathcal{X}=\{X\in\mathcal{A}\dvtx\overline{y}_A < \overline{y}_{X}\}$
(i.e., $X\in\mathcal{X}$ violates isotonicity with $A)$. The split in
$G$ causes the fit in nodes in $A$ to decrease. We will prove that when
the fits in $A$ decrease, there can be no groups below $A$ that become
violated by the new fits to $A$, that is, the decreased fits in $A$
cannot be such that $\mathcal{X}\neq\{\cdot\}$.

We first prove that $\mathcal{X}=\{\cdot\}$ by contradiction. Assume
$\mathcal{X}\neq\{\cdot\}$. Denote $i<k+1$ as the iteration at which the
last of the groups in $\mathcal{X}$, denoted~$D$, was split from $G$
and suppose at iteration $i$, $G$ was part of a larger group $H$ and~$D$ was part
of a larger group $F$. It is important to note that
$X\cap(F\cup H)=\{\cdot\}$ for all $X\in\mathcal{X}\setminus D$ at
iteration~$i$ because by assumption all groups in
$\mathcal{X}\setminus D$ were separated from $A$ before iteration~$i$.
Thus, at iteration~$i$, $D$ is the only group bordering $A$
that violates isotonicity.

Let $D_U$ denote the union of $D$ and all groups in $F$ that
majorize $D$. By construction, $D_U$ is a majorant in $F$. Hence,
$\overline{y}_{D_U} < \overline{y}_{F\cup H}$ by Algorithm
\ref{algIRnoregret} and $\overline{y}_A < \overline{y}_{D_U}$ by
definition since $\overline{y}_{D_U} > \overline{y}_D > \overline
{y}_A$. Also
by construction, any set $X\in H$ that minorizes $A$ has
$\overline{y}_{X} < \overline{y}_{A}$ (each set $X$ that minorizes
$A$ besides $D$ such that $\overline{y}_{X} < \overline{y}_{A}$ has
already been split from $A$). Hence, we can denote $A_L$ as the union
of $A$ and all groups in $H$ that minorize $A$ and we have
$\overline{y}_A
> \overline{y}_{A_L}$ and $A_L$ is a minorant in $H$. Since
$A_L\subseteq H$ at iteration $i$, we have
\[
\overline{y}_{F\cup H} < \overline{y}_{A_L} < \overline{y}_A <
\overline{y}_{D_U} <
\overline{y}_{F\cup H},
\]
which is a contradiction, and hence the assumption $\mathcal{X}\neq\{
\cdot\}$ is false. The first inequality is because the algorithm left $A_L$
in $H$ when $F$ was split from~$H$, and the remaining inequalities are
due to the above discussion. Hence, the split at iterations $k+1$ could
not have caused a break in isotonicity.

A similar argument can be made to show that the increased fit for
nodes in $B$ does not cause any isotonic violation. The proof is
hence completed by induction.
\end{pf}

With Theorem \ref{thisotonicsolutions}, the machinery for generating a
regularization path is complete. In Section \ref{scomplexity} we
describe the computational complexity for generating this path,
followed by a discussion of the statistical complexity of the solutions
along the path in Section \ref{sdegreesfreedom}.

\section{Complexity} \label{scomplexity}
We here show that the partitioning step in IRP can be solved
efficiently. The computational bottleneck of Algorithm \ref
{algIRnoregret} is solving the linear program (\ref{eqoptimalcutlp})
that iteratively partitions each group. The linear program~(\ref
{eqoptimalcutlp}) has a special structure
that can be taken advantage of in order to solve larger problems
faster. Indeed, the dual problem can be written as an optimization
problem called a network flow problem that is amenable to very
efficient algorithms, as noted by \citet{Spou2003} who recognize
the network flow problem as the \textit{maximal upper set problem}. We
note that our partition problem (\ref{eqoptimalcutlp}) is very similar
to the network flow problem solved in \citet{Chan2005} where
$z_i$ there represents the classification performance on node $i$.
We denote the complexity of solving the linear program (\ref
{eqoptimalcutlp}) by $T(m,n)$, where $m$ is the number of
constraints\vadjust{\goodbreak}
defined by $\mathcal{I}$ and~$n$ is the number of observations.
Various efficient algorithms for solving this problem exist, giving
complexities such as $T(m,n)=O(mn\log{n})$ [\citet{Slea1983}]
along with several algorithms giving $T(m,n)=O(n^3)$ [\citet
{Gali1980}]. Choosing the more efficient implementation depends on the
number of isotonicity constraints $m$ [e.g., $n^3\leq mn\log{n}$ for
the worst case $m=O(n^2)$]. A recent result by \citet{Stou2010}
shows how to represent an isotonic regression problem by an equivalent
problem where both the number of total observations and constraints are
of order $O(n\log^{d-1}{n})$, which greatly reduces the worst case of
$m=O(n^2)$ isotonicity constraints (i.e., by trading off a few
additional \textit{shadow} observations for a large reduction in the
number of constraints). Since at most $n$ partitions are made by IRP,
complexity is $O(n^4)$ using $T(m,n)=O(n^3)$ or reduced to $O(n^3\log
^{2d-1}{n})$ using the results of \citet{Stou2010}.

In practice, the complexity can be even better by accounting for the
fact that IRP solves a sequence of partitioning problems that are
decreasing in size (i.e., problems with fewer and fewer observations).
Each partition in Algorithm \ref{algIRnoregret} can be divided into
different proportions. We generically denote the bigger proportion in a
partition by $p \geq0.5$. Proposition \ref{propcomplexity} next gives
a~bound on the complexity of Algorithm \ref{algIRnoregret} for this
general case [assuming $T(m,n)=O(n^3)$], in terms of the maximal $p$
over all partitions.
\begin{proposition} \label{propcomplexity}
Let $p_{\max}\ge0.5$ be the greatest $p$ over all iterations of
Algorithm \ref{algIRnoregret} such that iteration $k$ partitions a
group of size $n_k$ into two groups of size $pn_k$ and $(1-p)n_k$.
Denote by $n$ the total number of observations. Then the complexity of
Algorithm \ref{algIRnoregret} is bounded by
%
\begin{equation} \label{eqtotalcomplexity}
O(n^3)\frac{1}{1-p_{\max}^2}.
\end{equation}
\end{proposition}

The proof, given in the \hyperref[app]{Appendix}, is based on the fact that the
sequence of IRPs partition problems are solved on smaller and smaller
groups of observations [i.e., while the first partition problem is
$O(n^3)$, the partition problems for the two created partitions are
$O(p^3n^3)$ and $O((1-p)^3n^3)$ for some $p$ where $0<p<1$]. Even at
$p_{\max}=0.99$, the constant $1/(1-p_{\max}^2)\approx50$, which is
very small when the number of observations is large. Thus, under
another reasonable assumption that $p_{\max}$ is bounded, we can
conclude that IRP is of practical complexity $O(n^3)/(1-p_{\max}^2)$.
Similar analysis using the results of \citet{Stou2010} lead to a
practical complexity of $O(n^2\log^{2d-1}{n})/\allowbreak(1-p_{\max})$.

Additional enhancements to the algorithm can be made in order to
solve even larger problems than done here. As noted above, the dual
problem to our optimal partition problem is a specially structured
network flow problem. \citet{Luss2010b} show that the network flow
problem [i.e., the dual to problem~(\ref{eqoptimalcutlp})] can be
decomposed and solved through a~sequence of smaller network flow
problems. This reduction of the optimal partition problem makes IRP
a practical tool for even larger data sets than experimented with
here; refer to \citet{Luss2010b} for details on the large-scale
decomposition which is beyond the scope of this paper.

\section{Degrees of freedom of isotonic regression and IRP}\label
{sdegreesfreedom}
The concept of degrees of freedom is commonly used in statistics to
measure the complexity of a model (or more accurately, a modeling
approach). This concept captures the amount of fitting the model
performs, as expressed by the optimism of the in-sample error
estimates, compared to out-of-sample predictive performance. Here we
briefly review the main ideas of this general approach, and then
apply them to isotonic regression and IRP.

Following \citet{Efron1986} and \citet{Hastie2001}, assume
the values $x_1,\ldots,x_n \in\mathbb R^d$ are fixed in advance (the
\textit{fixed-x}
assumption), and that the model gets one vector of observations
\mbox{$\by\!=\!(y_1,\ldots,y_n)^\tr\!\in\!\mathbb{R}^n$} for training, drawn according
to $P(y|x)$ at the $n$ data points. Denote by~$\byn$ another
independent vector drawn according to the same distribution.~$\by$
is used for training a model $\hat{f}(x)$ and generating
predictions $\hat{y}_i = \hat{f}(x_i)$ at the~$n$ data points.

We define the \textit{in-sample} mean squared error,
\[
\mathrm{MRSS} = \frac{1}{n} \| \by-\hat{\by}\|_2^2,
\]
and compare it to the expected error the same model incurs on the
new, independent copy, denoted in \citet{Hastie2001} by~$\ei$,
\[
\ei= \frac{1}{n} \mathbb E_{\byn} \| \byn-\hat{\by}\|_2^2.
\]
The difference between the two is the \textit{optimism} of the in-sample
prediction. As \citet{Efron1986} and others have shown, the expected
optimism in MRSS is
%
\begin{equation} \label{op}
E_{\by,\byn} (\ei- \mathrm{MRSS}) = \frac{2}{n} \sum_i
\operatorname{cov}(y_i,\hy_i).
\end{equation}
For linear regression with homoskedastic errors with variance $\sigma^2$,
it is easy to show that (\ref{op}) is equal to $\frac{2}{n} \,d\sigma
^2$, where $d$ is the number
of regressors, hence the degrees of freedom. This naturally leads to
defining the \textit{equivalent degrees of
freedom} of a modeling approach as
%
\begin{equation} \label{eqdf}
\df=\sum_i{\operatorname{cov}(y_i,\hy_i)}/\sigma^2.
\end{equation}

In nonparametric models, one usually cannot calculate the actual
degrees of freedom of a modeling approach, but it is often easier to
generate \textit{unbiased estimates}\vadjust{\goodbreak} $\hat{\df}$ of $\df$ using Stein's
lemma [\citet{Stein81}]. \citet{Meyer2000} demonstrate the
applicability of this theory in shape-restricted nonparametric
regression. Specifically, their Proposition 2, adapted to our
notation, implies that if we assume the homoskedastic case $\operatorname{var}(y_i)
= \sigma^2$ for all $i$, then the unbiased estimator $\hat{\df}$ for
degrees of freedom in isotonic regression is the expected number of pieces
$D$ in the solution $\bhy$ to (\ref{eqIR}), that is,
\[
\mE(D)=\sum_i{\operatorname{cov}(y_i,\hy_i)}/\sigma^2.
\]

Considering the IRP algorithm, this puts us in the interesting
situation where the number of steps the algorithm takes until it
terminates in the globally optimal isotonic solution is equal to the degrees
of freedom estimator of this global solution (minus one, since we
start with one piece). One might thus be inclined to assume that
each iteration of Algorithm \ref{algIRnoregret} adds \textit{about}
one degree of freedom, that is, performs approximately the same amount
of fitting in every iteration. A similar idea is represented by the
degrees of freedom calculation of \citet{Sche1997} in their
reduced monotonic regression algorithm (which starts from the
complete isotonic fit and eliminates pieces).

On more careful consideration, however, it is obvious that this idea is
incorrect since the first iteration of IRP finds an optimal cut
in the very large space of all possible multivariate isotonic cuts.
For comparison, a single deep split in a regression tree has been
estimated to
consume three or more degrees of freedom [\citet{Ye1998}], and
the space of possible splits
in initial IRP iterations is much larger than that of a regression tree
since IRP splits are not limited to being axis-oriented.
Thus, intuitively, the first iteration is expected to use much more
than one degree of
freedom (the equivalent of fitting one coefficient to a \textit{fixed},
\textit{pre-determined} regressor). This effect should be exacerbated
as the
dimension $d$ of $x$ increases since the size of the search space for
isotonic cuts increases with it. It also inevitably implies that the
latter iterations of the IRP algorithm should perform less
(ultimately much less) fitting than the equivalent of one degree of
freedom in every iteration in order to be consistent with the unbiasedness
of $\hat{\df}=D$ as an estimator of $\df$.

\begin{figure}

\includegraphics{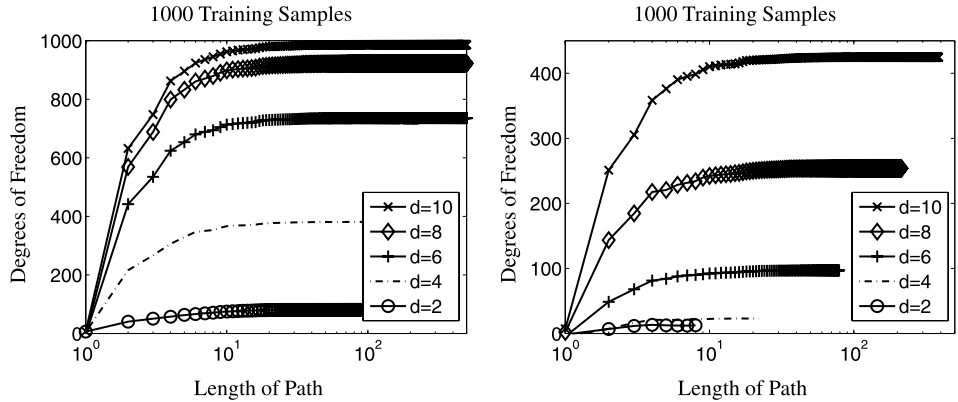}

\caption{Evolution of degrees of freedom for IRP as model complexity
increases. Both models use $y_i=\sum_j{x^2_{ij}}+\mathcal{N}(0,10)$.
Simulation (left) uses $x_{ij}\sim\mathcal{U}[1,2]$ and simulation
(right) uses $x_{ij}\in\{0,1,2\}$ with probabilities $\{1/3,1/3,1/3\}
$. Each path is the mean over 50 trials with 1,000 training samples.}
\label{figDFn1000}
\end{figure}

Here we demonstrate empirically that this is indeed the case. We
simulate data from a simple additive model
%
\begin{eqnarray}
\label{eqxdf}
x_{ij}&\sim&\mathcal{U}[1,2] \quad\mbox{i.i.d.},\\
\label{eqydf}
y_i&=&\sum_j {x_{ij}^2}+\mathcal{N}(0,10),
\end{eqnarray}
where $x_{ij}$ is dimension $j$ of the observation $i$. We can generate
one fixed copy of data using (\ref{eqxdf}), and then repeatedly
generate observations using~(\ref{eqydf}), apply IRP, and empirically
estimate $\df$ as defined by
(\ref{eqdf}) for every iteration of IRP. Figure\vadjust{\goodbreak} \ref{figDFn1000}(left)
shows how $\df$ evolves in this model as the IRP iterations proceed,
%
for increasing dimensions of $x$. The covariance in~(\ref{eqdf})
was estimated by drawing values $X = (x_1,\ldots,x_{1\mbox{,}000})$ according
to the model~(\ref{eqxdf}), fixing them, and repeatedly drawing $1\mbox{,}000$
independent
copies of~$\by|X$ according to (\ref{eqydf}), and applying IRP on
each one. 200 simulations were run with one drawing of $X$ and the results
were averaged. As expected, we see that the number of pieces (hence
degrees of freedom) in the final isotonic regression increases with
the dimension, as does the rate in which the number of degrees of
freedom increases in the initial steps of IRP.

\begin{figure}

\includegraphics{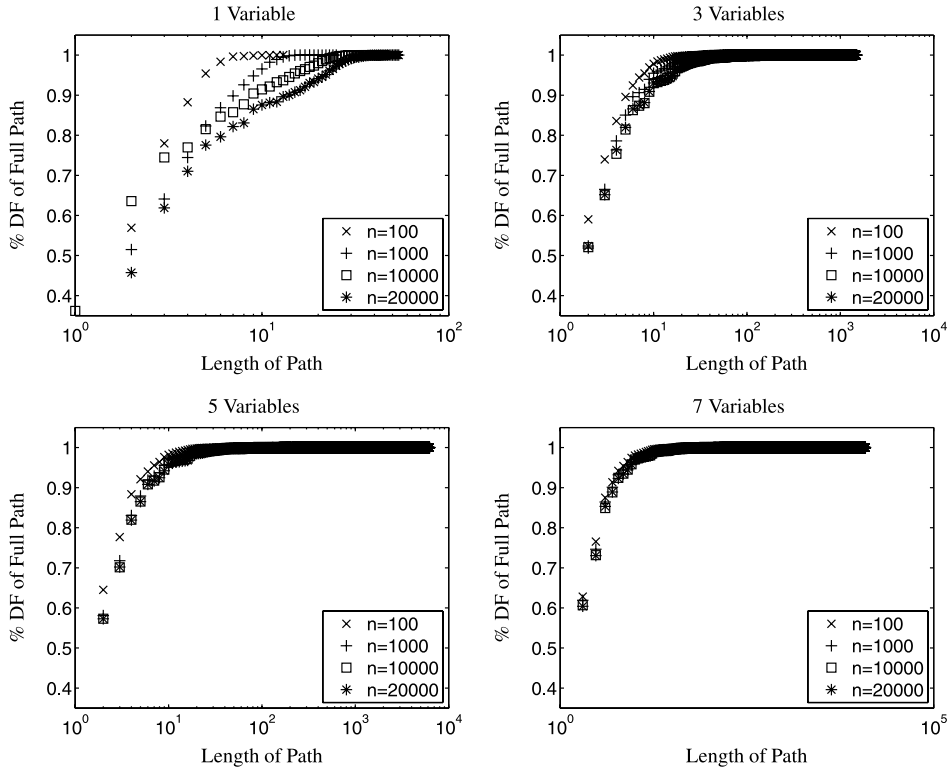}

\caption{Percentage of degrees of freedom relative to the full path
(i.e., number of partitions in the optimal solution)
as model complexity increases. Simulations use $\bx_{ij}\sim\mathcal
{U}[1,2]$ with
$y_i=\sum_j{x^2_{ij}}+\mathcal{N}(0,10)$.
Each path is the mean over 200 trials. Only the first 500 partitions of
the paths are displayed in order to make the MSE of the earlier IRP
iterations visually clearer.} \label{figDFpercentage}
\end{figure}

In order to emphasize this dependence of the degrees of freedom in
initial iterations on
the dimension, as well as on the number of observations, Figure \ref
{figDFpercentage} presents the evolution of the percentage of
the total isotonic regression degrees of freedom along the path (i.e.,
number of degrees of freedom relative to the number of partitions of
%
the final model) as a function of both the dimension and the amount of
data used. As
expected, increasing the dimension radically increases the portion
of the fitting in the first steps, while increasing the amount of
data decreases this portion (since the overall isotonic fit is
generally more complex in these situations). It should be noted that
for many of the situations examined, IRP performs more than half of
the total isotonic fit, as measured by degrees of freedom, in its
first iteration! In dimension 7, even at $n=20\mbox{,}000$
observations, almost $60\%$ of the total fitting is associated with
the first iteration. Thus, these simulations clearly demonstrate the
nature and limitations of IRPs regularization behavior: the IRP
path contains models that are regularized isotonic
models compared to the global solution, but IRPs ability to control model
complexity is limited by the concentration of most of the fitting in
the initial iterations, especially in higher dimension.

As mentioned in the
\hyperref[sintroduction]{Introduction}, an area that combines applications
where the isotonic assumptions are reasonable (i.e., low bias) and
the overfitting may be of less concern (i.e., variance can be
controlled) is in genetics, specifically in modeling gene--gene
interactions in phenotype(y)-genotype(X) relationships
[\citet{Cordell2009}]. The key observation here is that genotypes are
ternary ($x_{ij} \in\{0,1,2\}$ copies of the ``risk'' allele).
Thus, each dimension of the predictor space $\mathcal X$ can take only
one of three possible observed values in the data. Intuitively, it
is clear that this would significantly reduce the space of possible
isotonic splits in IRP, and hence reduce the amount of fitting. To
demonstrate this empirically, Figure \ref{figDFn1000}(right)
displays an experiment with the same setup, where instead of drawing
the $x$ values from a multivariate uniform, they are drawn
independently from $\{0,1,2\}$ with equal probabilities and we use
the same model. Figure \ref{figDFn1000} demonstrates that both the
globally optimal isotonic regression and, especially, the first IRP
iterations perform much less overall fitting in the ternary case
versus the continuous case, as measured by equivalent degrees of
freedom. For example, in six dimensions, the continuous case
requires almost seven times as many degrees of freedom than in the
ternary case to fit the model. However, relative to the final model,
a large percentage of the fitting still takes place in the initial
iterations.

\section{Performance evaluation} \label{sperformance}
We demonstrate here the usefulness of isotonic regression on
simulation and real data. For each experiment, IRP is run on the
training data and produces a path of isotonic models. Each model is
used for prediction on the test data and the root mean squared error
(RMSE) is recorded. This generates paths of root mean squared errors
over the different isotonic models and is illustrated in the figures
below. In each table, we record the minimum RMSE along these paths
(\textit{IRP min RMSE}), along with how many partitions were made to
generate this minimum RMSE (\textit{IRP min path}), and the number of
partitions in the global isotonic solution (\textit{IRP path length}).
IRP, as well as optimal isotonic regression, results are compared to
running a least squares regression on the training data and
predicting on the testing data with the resulting linear model
(corresponding RMSE is called \textit{LS RMSE}), and to the
performance of the global isotonic regression solution
(\textit{Isotonic regression RMSE}). Because we are interested in
examining the behavior of the entire IRP path for use in selecting the
optimal tuning parameter, and to avoid a significant increase in
running time, we do not employ cross-validation for selecting the
best stopping point (number of iterations), but use test sets for
this. In practical application, cross-validation would be the
appropriate approach for selecting the best model for prediction.

\subsection{Simulations}
\label{sssimulations} We first illustrate isotonic regression on
simulated data with different distributions. For the first three
experiments, the $i$th observation's regressors are distributed
as $x_{ij}\sim\mathcal{U}[0,5]$, $x_{ij}\sim\mathcal{U}[0,3]$, and
$x_{ij}\sim\mathcal{U}[0,2]$, respectively, and in all cases, all
$x_{ij}$ are i.i.d. Responses $y_i$ for the three simulations are
generated as
\begin{eqnarray*}
\mbox{(1)\quad}y_i=\biggl(\sum_j{x^2_{ij}}\biggr)+\mathcal{N}(0,4d^2),
\qquad\mbox{(2)\quad}y_i=\Biggl(\prod_{j=1}^d{x_{ij}}\Biggr)+\mathcal{N}(0,d^2)
\end{eqnarray*}
and
\[
\mbox{(3)\quad}y_i=2^{\sum_j{x_{ij}}}+\mathcal{N}(0,d^2),
\]
respectively, where $d$ is the dimension. The first simulation
represents an additive (nonlinear) model and the other two
simulations are ``super-additive'' models (i.e., represent strong
positive interactions which are hard to approximate with
additive
effects). This allows us to examine the performance of IRP and
isotonic regression in a variety of relevant situations.\vadjust{\goodbreak}

The last two experiments are ternary. The $i$th observation's
regressors are distributed as $x_{ij}\in\{0,1,2\}$ with
probabilities $\{0.7,0.2,0.1\}$ and $\{1/3,1/3,\allowbreak1/3\}$ for the fourth
and fifth experiments, respectively. The fourth model is
subadditive while the fifth model is superadditive; specifically,
they are
\[
\mbox{(4)\quad}y_i=\biggl(\sum_j{x_{ij}}\biggr)^{1/4}+\mathcal{N}(0,d^2/10)
\quad\mbox{and}\quad \mbox{(5)\quad}y_i=\biggl(\prod_j{x_{ij}}\biggr)+\mathcal{N}(0,d^2).
\]
Model 4 is chosen to simulate ``subadditive'' genetic interactions
as discovered by \citet{Shao2008}, where having one risk variant is
sufficient to attain most of the effect, and additional variants
have little additional influence. Model 5 represents the
``superadditive'' model, where variants exacerbate each other's
effect, as often speculated to be the case in human disease. Note
that, for each of 50 simulations, 12,000 training and 3,000 testing
points were randomly generated and statistics computed (all tests
are out-of-sample).

%
\begin{figure}

\includegraphics{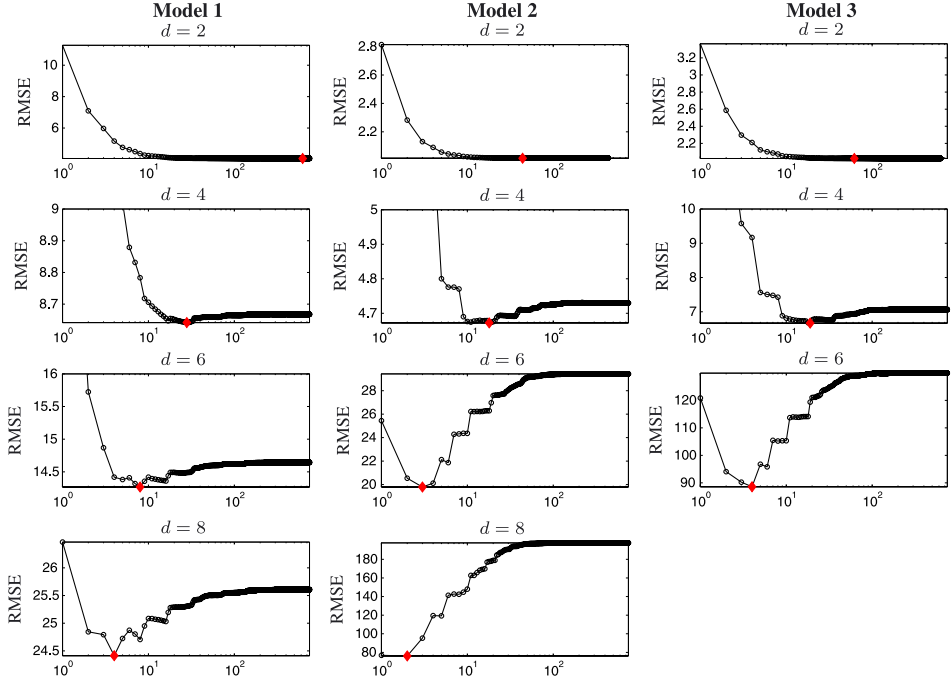}

\caption{Root mean squared error (RMSE) for out-of-sample
predictions of simulations with different dimensions $d$. The
x-axis in each figure corresponds to the number of partitions made
by IRP, that is, the curves show how the RMSE of test data varies as the
IRP algorithm progresses. Model 1 uses the function
$y_i=(\sum_j{x^2_{ij}})+\mathcal{N}(0,4d^2)$ with
$x_{ij}\sim\mathcal{U}[0,5]$. Model 2 uses\vspace*{-1pt} the function
$y_i=(\prod_j{x_{ij}})+\mathcal{N}(0,d^2)$ with
$x_{ij}\sim\mathcal{U}[0,3]$. Model 3 uses the function
$y_i=2^{\sum_j{x_{ij}}}+\mathcal{N}(0,d^2)$ with
$x_{ij}\sim\mathcal{U}[0,2]$. Fifty simulations were run with 12,000
training and 3,000 testing points. Only the first 750 partitions of
the paths are displayed in order to make the RMSE of the earlier IRP
iterations visually clearer.} \label{figsimulations}
\end{figure}

Figure \ref{figsimulations} demonstrates testing error for IRP over
the regularized path of isotonic solutions for the first three
experiments (with continuous covariates). The main
observation here is that as the dimension increases, the effect of
overfitting of the standard (nonregularized) isotonic regression
becomes more significant and causes the skewed $U$-shaped pattern
across the IRP path, where the minimum prediction RMSE is obtained
earlier in the path. This is the effect we alluded to in the
\hyperref[sintroduction]{Introduction} and it stems from the
limitations of the isotonicity
constraints in controlling model complexity in high dimensions.

%
\begin{table}
\caption{Statistics for simulations generated by the three different
models as labeled above. A path of root mean squared errors (RMSE)
for each model along the regularization path was computed. {IRP
min RMSE} refers to the minimum RMSE along these paths. Values for
RMSE are given along with a conservative 95\% confidence interval.
{LS RMSE} is the RMSE from using least squares regressions.
{IRP min path} is the number of partitions made to generate the
minimum RMSE and IRP path length is the number of partitions in the
global isotonic solution. Bolded RMSE values for IRP and isotonic
regression indicate that they are lower than the RMSE of the least
squares regression with 95\% confidence}
\label{tablesimulationstats}
\begin{tabular*}{\tablewidth}{@{\extracolsep{4in minus 4in}}lrrrd{3.0}c@{}}
\hline
\multicolumn{1}{@{}l}{\textbf{Number of}} &
\multicolumn{1}{c}{\textbf{IRP min}} &
\multicolumn{1}{c}{\textbf{Isotonic regression}\hspace*{-6pt}}
& \multicolumn{1}{c}{\textbf{LS}} & \multicolumn{1}{c}{\textbf{IRP min}} &
\multicolumn{1}{c@{}}{\textbf{IRP path}} \\
\multicolumn{1}{@{}l}{\textbf{variables}}
& \multicolumn{1}{c}{\textbf{RMSE}} & \multicolumn{1}{c}{\textbf{RMSE}\hspace*{-6pt}}
& \multicolumn{1}{c}{\textbf{RMSE}}
& \multicolumn{1}{c}{\textbf{path}} & \multicolumn{1}{c@{}}{\textbf{length}} \\
\hline
\multicolumn{6}{c}{Model 1:
$y_i=(\sum_j{x^2_{ij}})+\mathcal{N}(0,4d^2)$ with
$x_{ij}\sim\mathcal{U}[0,5]$} \\ [4pt]
2&\textbf{4.06 ($\pm$0.02)}&\textbf{4.06 ($\pm$0.02)}\hphantom{0}&4.80 ($\pm
$0.01)&625&\hphantom{00,}999\\
4&\textbf{8.64 ($\pm$0.03)}&\textbf{8.67 ($\pm$0.04)}\hphantom{0}&8.83 ($\pm
$0.03)&28&\hphantom{0}4,861\\
6&14.27 ($\pm$0.06)&14.64 ($\pm$0.07)\hphantom{0}&12.87 ($\pm$0.04)&8&\hphantom{0}8,520\\
8&24.41 ($\pm$0.11)&25.61 ($\pm$0.12)\hphantom{0}&16.89 ($\pm$0.05)&4&10,604\\
[6pt]
\multicolumn{6}{c}{Model 2:
$y_i=(\prod_j{x_{ij}})+\mathcal{N}(0,d^2)$ with
$x_{ij}\sim\mathcal{U}[0,3]$} \\
[4pt]
2&\textbf{2.01 ($\pm$0.01)}&\textbf{2.01 ($\pm$0.01)}\hphantom{0}&2.13 ($\pm
$0.01)&44&\hphantom{00,}437\\
4&\textbf{4.67 ($\pm$0.03)}&\textbf{4.73 ($\pm$0.04)}\hphantom{0}&6.07 ($\pm
$0.03)&18&\hphantom{0}3,711\\
6&19.79 ($\pm$0.24)&29.43 ($\pm$0.97)\hphantom{0}&19.62 ($\pm$0.29)&3&\hphantom{0}7,685\\
8&76.23 ($\pm$2.08)&197.71 ($\pm$16.49)&65.40 ($\pm$2.24)&2&10,153\\
[6pt]
\multicolumn{6}{c}{Model 3:
$y_i=2^{\sum_j{x_{ij}}}+\mathcal{N}(0,d^2)$ with
$x_{ij}\sim\mathcal{U}[0,2]$} \\
[4pt]
2&\textbf{2.02 ($\pm$0.01)}&\textbf{2.02 ($\pm$0.01)}\hphantom{0}&2.17 ($\pm
$0.01)&62&\hphantom{00,}628\\
4&\textbf{6.67 ($\pm$0.08)}&\textbf{7.06 ($\pm$0.14)}\hphantom{0}&10.10 ($\pm
$0.09)&19&\hphantom{0}7,558\\
6&88.59 ($\pm$1.13)&129.91 ($\pm$4.13)\hphantom{0}&70.77 ($\pm$1.21)&4&11,787\\
[6pt]
\multicolumn{6}{c}{Model 4:
$y_i=(\sum_j{x_{ij}})^{1/4}+\mathcal{N}(0,d^2/10)$ with
$x_{ij}\in\{0,1,2\}$ with probabilities $\{0.7,0.2,0.1\}$} \\
[4pt]
2&\textbf{0.63 ($\pm$0.00)}&\textbf{0.63 ($\pm$0.00)}\hphantom{0}&0.67 ($\pm
$0.00)&8&\hphantom{0000,}8\\
4&\textbf{1.27 ($\pm$0.01)}&\textbf{1.27 ($\pm$0.01)}\hphantom{0}&1.29 ($\pm
$0.01)&9&\hphantom{000,}30\\
6&1.91 ($\pm$0.01)&1.91 ($\pm$0.01)\hphantom{0}&1.92 ($\pm$0.01)&4&\hphantom{000,}85\\
8&2.55 ($\pm$0.01)&2.56 ($\pm$0.01)\hphantom{0}&2.54 ($\pm$0.01)&5&\hphantom{00,}267\\
[6pt]
\multicolumn{6}{c}{Model 5:
$y_i=(\prod_j{x_{ij}})+\mathcal{N}(0,d^2)$ with $x_{ij}\in\{0,1,2\}$
with probabilities $\{1/3,1/3,1/3\}$} \\
[4pt]
2&\textbf{2.00 ($\pm$0.01)}&\textbf{2.00 ($\pm$0.01)}\hphantom{0}&2.11 ($\pm
$0.01)&4&\hphantom{0000,}5\\
4&\textbf{4.00 ($\pm$0.02)}&\textbf{4.00 ($\pm$0.02)}\hphantom{0}&4.47 ($\pm
$0.02)&6&\hphantom{000,}25\\
6&\textbf{6.04 ($\pm$0.02)}&\textbf{6.04 ($\pm$0.02)}\hphantom{0}&7.27 ($\pm
$0.05)&87&\hphantom{00,}103\\
8&\textbf{8.30 ($\pm$0.04)}&\textbf{8.30 ($\pm$0.04)}\hphantom{0}&10.90 ($\pm
$0.21)&67&\hphantom{00,}430\\
\hline
\end{tabular*}
\end{table}

Table \ref{tablesimulationstats} displays certain statistics on
all five simulations as well as a comparison to the results of a
least squares regression. We first discuss the case of continuous
covariates (first three models). In lower dimensions standard
isotonic regression performs well, and regularization through IRP
offers no gain (this is seen in dimension $d=2$ in all three
examples). Here, isotonic regression controls bias by accommodating
the nonlinearities in the true model and significantly
outperforms least squares regression. As the number of covariates
increases, regularization through IRP becomes necessary to control
variance, and the optimal performance is obtained earlier in the IRP
path (dimension $d=4$ in our examples). When $d$ increases further,
however, IRP also becomes inefficient at controlling variance, and
linear regression dominates. This effect can be traced back to the
large amount of fitting performed by IRP already in its initial
iterations, as demonstrated in the previous section.

With respect to the models with ternary covariates, isotonic
regression outperforms the simple linear regression, however, the IRP
path does not statistically improve performance. For the
subadditive model (model 4), performance is better for dimensions 2
and 4, after which again IRP is unsuccessful at controlling
variance. However, in the superadditive model (model 5), IRP
dominates for all dimensions. This is not surprising, since
superadditive models are more extreme in their deviations from
additivity, making the flexibility of IRP more critical.

Thus, our simulations confirm that isotonic regression performs well
in low dimensions, but requires a lot of data in order to learn good
nonlinear models in higher dimensions. In intermediate dimensions,
IRP can offer a~compromise between fitting flexible isotonic models
and controlling model complexity, resulting in useful prediction
models.

\subsection{Modeling MPG of automobiles} \label{realdataexamples}
We next illustrate the performance of IRP when predicting the
miles-per-gallon of a list of 392 automobiles manufactured between
1970 and 1982 using seven variables [\citet{Asun2010}]. Seven
regressions are performed in dimensions one through seven, where the
variables chosen are from the following order: origin (American,
European or Japanese), model year, number of cylinders,
acceleration, displacement, horsepower, and weight. The order of
the variables was determined in order of the magnitude of
coefficients from a least squares linear regression on all
variables. Origin, surprisingly, had the largest magnitude, and in
giving discrete variables 1, 2, 3 to the respective origins, there
actually is a monotonic trend in origin (i.e., American cars are
least fuel efficient, followed by European cars, with the Japanese
being most efficient). While we include origin as a variable here,
we note that similar performance for IRP was achieved without origin
in an independent experiment.

%
\begin{table}
\caption{Statistics for auto mpg data. Miles-per-gallon is
regressed on seven potential variables: origin (American,
European or Japanese), model year, number of cylinders,
acceleration, displacement, horsepower, and weight. Row $k$ uses the
first $k$ variables from this list in the regression. A path of root
mean squared errors for each model along the regularization path was
computed. Bold demonstrates statistical significance of either IRP
or isotonic regression over a least squares regression with 95\%
confidence, as determined by a paired t-test using 392 observed
squared losses obtained from leave-one-out cross-validation}
\label{tableautostats}
\begin{tabular*}{\tablewidth}{@{\extracolsep{\fill}}lccccc@{}}
\hline
\textbf{Number of}& \textbf{IRP min} & \textbf{Isotonic regression}
& \textbf{LS} & \textbf{IRP min} & \textbf{IRP path}\\
\textbf{variables} & \textbf{RMSE} & \textbf{RMSE} & \textbf{RMSE}
& \textbf{path} & \textbf{length} \\
\hline
1&6.46 ($\pm$0.60)&6.50 ($\pm$0.43)&6.46 ($\pm$0.49)&\hphantom{0}9&\hphantom{0}17\\
2&\textbf{4.91 ($\pm$0.82)}&\textbf{4.95 ($\pm$0.37)}&5.24 ($\pm
$0.37)&\hphantom{0}7&\hphantom{0}26\\
3&\textbf{3.73 ($\pm$1.17)}&\textbf{3.76 ($\pm$0.36)}&4.02 ($\pm
$0.38)&\hphantom{0}9&\hphantom{0}37\\
4&3.83 ($\pm$1.13)&3.91 ($\pm$0.37)&4.04 ($\pm$0.38)&\hphantom{0}7&\hphantom{0}62\\
5&\textbf{3.32 ($\pm$1.41)}&\textbf{3.36 ($\pm$0.38)}&3.92 ($\pm
$0.40)&15&109\\
6&\textbf{3.28 ($\pm$1.43)}&3.37 ($\pm$0.37)&3.77 ($\pm$0.38)&15&114\\
7&3.29 ($\pm$1.43)&3.36 ($\pm$0.37)&3.37 ($\pm$0.33)&\hphantom{0}8&128\\
\hline
\end{tabular*}
\end{table}

\begin{figure}

\includegraphics{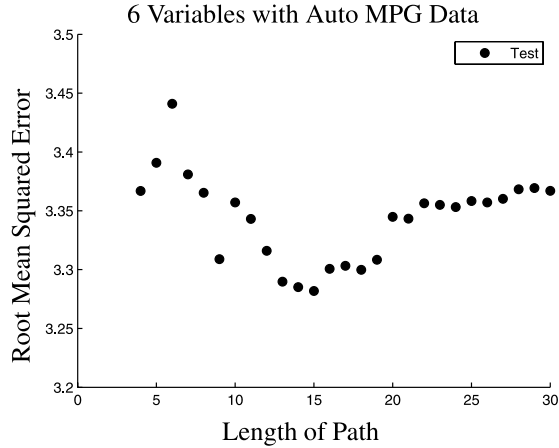}

\caption{Root mean squared error (RMSE) for auto data with six
variables using IRP illustrates that overfitting occurs after 15
iterations of IRP, that is, RMSE decreases until~15 partitions are
made, after which RMSE begins to increase. The figure displays only the
first 30 partitions so that the U-shape is clear.} \label{figauto6vars}
\end{figure}

Since the data set is rather small, we perform leave-one-out
cross-validation (i.e., the data is divided into training and testing
sets of 391 and 1 instances, respectively, so that each instance is
used out-of-sample once). Table \ref{tableautostats} displays
certain statistics on the IRP, and isotonic regression, performance
as well as a comparison to the results of a least squares
regression. Figure \ref{figauto6vars} displays MSE on
out-of-sample data for IRP over the regularized path of isotonic
solutions for a regression with six variables, exemplifying that
overfitting occurs after 15 iterations of IRP (seen by the $U$-shaped
curve with minimum at 15 iterations).

\section{The search for gene--gene interactions} \label{sepistasis}
As mentioned in Section \ref{sintroduction}, the search for
gene--gene interactions (epistasis) is a major endeavor in the
genetics community, with the goal to identify the mechanisms that
connect genotypes and phenotypes of interest, including height and
disease.

It is generally acknowledged that the search has so far yielded very
limited actual findings of major and impactful epistasis in human
phenotypes [\citet{Cordell2009}]. Where significant interactions have
been found, like recently by \citet{Zhang2011}, these were often
limited to close-by regions in the genome, where statistical and
biological interactions are hard to differentiate (because mutation
distributions are correlated due to linkage disequilibrium). If our
interest is focused on biological interactions (e.g., two mutations
influencing each other's causative effects on the phenotype), we
should be particularly interested in finding epistasis between
noncorrelated mutations.

The limited findings in the literature may be due to two distinct
reasons: first, the difficulty of searching through the space of all
possible combinations of mutations. For example, in typical genome
wide association studies (GWAS), one genotypes hundreds of thousands
of single nucleotide polymorphisms (SNPs), yielding billions of
potential two-way interactions, and much larger numbers of higher
order interactions. Heuristics for searching, like limiting the search
to interactions between SNPs that are individually associated with
the phenotype, may miss combinations with weak main effects but
strong interactions. Second, the limitation to simple statistical
models like chi-square tests and logistic regression with explicit
interactions may not allow identifying complex high-order
interactions even within the searched space. IRP offers a remedy to
this second concern, in allowing the modeling of complex
interactions subject to isotonicity only. As our simulations have
shown, good performance on ternary data is expected up to dimension
six and even beyond when a truly strong epistatic signal is present.

%

To empirically examine epistasis in human disease using IRP, we
obtained data from the Welcome Trust Case Control Consortium
[\citet{WTCCC2007}], encompassing around 2,000 cases for each of
three diseases: Crohn's disease (CD), Type-I diabetes (T1D), and
Type-II diabetes (T2D). These are compared to 3,000 healthy controls.
All of these samples were genotyped for around 500,000 SNPs, and each
phenotype (disease) was analyzed for association between each SNP
status and case-control status using chi-square and logistic
regression. For each disease, between five and nine significant SNPs
were discovered after careful multiple comparison corrections
[\citet{WTCCC2007}].

We discuss in detail our modeling of the CD data set. It comprises
2,000 cases and 3,000 controls. Our covariates include nine unlinked
SNPs from seven different chromosomes, which were discovered as
significant after a~multiple comparison correction in the WTCCC GWAS
and at least one other study [\citet{GWAS2011}]. These are genotypes,
that is, ternary variables in $\{0,1,2\}$. This is a classification
problem with binary response (i.e., $y_i\in\{0,1\}$ for control vs
case), and we thus model the risk of Crohn's disease with an
isotonic logistic regression, rather than an $l_2$ isotonic
regression as we have done for continuous regressions. Specifically,
we fit isotonic models by maximizing the in-sample logistic
log-likelihood rather than the sum of squares:
%
\begin{equation} \label{eqlogist}\qquad
\max\Biggl\{ \sum_{i=1}^n{y_i \log{(\hat{p}_i)} +
(1-y_i)\log{(1-\hat{p_i})}} \dvtx\hat{p}_i\leq\hat{p}_j \mbox{ for
all } (i,j)\in\mathcal{I}\Biggr\}.
\end{equation}
As explicated by \citet{Bacc1989} and \citet{Auh2006}, the solution
to (\ref{eqlogist}) is identical to the solution from solving the
$l_2$ squared loss isotonic regression problem (\ref{eqIR}) when the
values $y_i \in\{0,1\}$. Since we solve $l_2$ isotonic regression
(\ref{eqIR}) using the IRP algorithm, we can use it to solve
the isotonic logistic regression problem.

To evaluate isotonic model performance, we compare cross-validated
area under the ROC curve (AUC) for each model in the IRP path to
that of the linear logistic regression model with the same
covariates that assume no interaction between the SNPs. We employ
150-fold cross-validation, and use the approach of \citet{Delong1988}
to calculate $p$-values on the holdout AUC difference.

\begin{figure}

\includegraphics{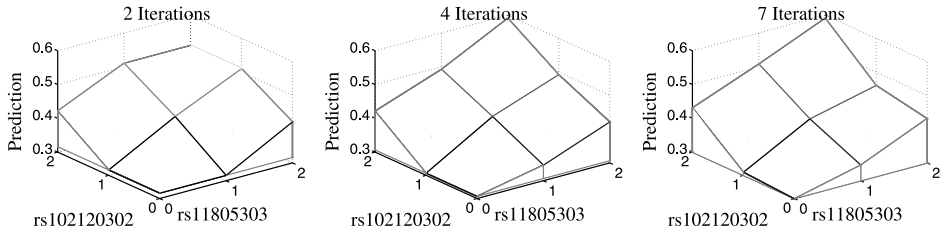}

\caption{Illustration of IRP with logistic log-likelihood loss
function for modeling Crohn's disease, as a function of two SNPs:
rs11805303 on chromosome 1 and rs102120302 on chromosome 2. Models
after iterations 2, 4 and 7 of IRP are shown. The model at
iteration~7 gives the best predictive performance along the path.}
\label{figcrohns}
\end{figure}

Figure \ref{figcrohns} illustrates the approach on a subset of two
SNPs. We can see the results of IRP after 2, 4 and 7 iterations (the
complete path has 8 iterations). The model at iteration 7 gives AUC
of 0.5562, compared to 0.5501 for logistic regression, yielding a
$p$-value of 0.0282 (which is only mildly indicative of a~possible
interaction given multiple comparison issues, as we hand-picked the
two mutations and the number of iterations). The fit at iteration 7
seems to support a super-additive interaction: presence of two
copies each from both SNPs confers a jump in CD risk.

%

We applied IRP on all subsets of three and four SNPs, and also the
entire set of nine SNPs. Several models yielded AUC improvements
over logistic regression which were significant at the nominal 0.05
level (like the two-variable model above), but none would withstand
a multiple comparison correction. With all SNPs, the full path
length was 96 iterations. The minimum model\vadjust{\goodbreak} AUC for isotonic
logistic regression was 0.6457 at 55 iterations and for logistic
regression was 0.6449. Because AUC was highly variable between
folds for this high-dimensional model, this difference was
insignificant: $p$-value was 0.4. The final model had an AUC of 0.6456
and the $p$-value was~0.42.

Our results on the T1D and T2D data sets were qualitatively similar:
no high order interactions of significance were identified by IRP.
As the IRP approach possesses the flexibility and power to identify
such interactions if they are strong enough (as illustrated in our
simulations), we can conclude that the strong univariate signals
identified by the WTCCC studies do not yield significant higher
order interactions. This is of course in line with previous findings
[\citet{Cordell2009}, \citet{Emily2009}], but now also confirmed by our more
sophisticated and flexible approach.

\section{Discussion and extensions} \label{sdiscussion}
The IRP algorithm offers solutions to both the statistical and
computational difficulties of isotonic regression. Algorithmically,
IRP solves (\ref{eqIR}) as a sequence of easier binary partitioning
problems that are efficiently solved using network flow algorithms.
From the statistical perspective, IRP generates a path of isotonic
models, each defining a partitioning of the space $\mathcal X$ into
isotonic regions. The averages of observations in these regions
comply with the isotonicity constraints
(Theorem \ref{thisotonicsolutions}). In this view, IRP provides
isotonic solutions along its path that are regularized versions of
the globally optimal isotonic regression solution.

Our discussion so far has focused on using the sum of squares loss
function in (\ref{eqIR}) for fitting ``standard'' isotonic
regression subject to squared error loss as well the logistic
log-likelihood which we noted is an identical problem. A well-known
result of \citet{Barl1972} implies that the solution of a whole
variety of loss functions subject to isotonicity constraints can be
obtained by solving standard isotonic regression,\vspace*{1pt} as long as the
loss can be written as minimizing $\sum_{i=1}^n w_i (\Psi(z_i) - x_i
z_i)^2$ in $z\in\reals^n$ for some convex differentiable $\Psi$ and
some data-dependent values $x$ and weights $w$. These results imply
that many other loss functions subject to isotonicity constraints
can optimally be solved by IRP via a reformulation to a problem of
the form (\ref{eqIR}). \citet{Barl1972} note that such
transformations can be applied to many maximum likelihood estimation
problems (subject to isotonicity constraints), including Bernoulli
(as described in Section \ref{sepistasis}), multinomial, poisson
and gamma distributed problems. We plan to investigate the
applicability of the resulting regularization algorithms in future
work.

We can now also formalize the connection between IRP and the
well-known work of \citet{Maxw1985} [and similarly
\citet{Round1986}] that was mentioned in the \hyperref
[sintroduction]{Introduction}.
\citet{Maxw1985} solved an operations research problem (related to
scheduling reorder intervals for a production system) by reducing it
to the optimization of a convex objective subject to isotonicity
constraints. In our notation, their objective (i.e., loss function)
is $\sum_{i=1}^n{(c_i/\hat{y}_i +b_i \hat{y}_i)}$, where $c_i$,
$b_i$ are data-dependent nonnegative constants determined by their
problem formulation. To apply the theory of \citet{Barl1972}, we
reformulate their problem as minimizing $\sum_{i=1}^n{c_i(z_i -
(-b_i/c_i))^2}$ in $z\in\reals^n$, that is, a standard weighted isotonic
regression, and recovering $\hat{y}_i^*=\sqrt{-z_i^*}$ (note that
the isotonic regression fits nonpositive observations $-b_i/c_i$).
Indeed, the algorithm of \citet{Maxw1985} is completely equivalent
to applying IRP on this modified problem! It should be emphasized,
however, that \citet{Maxw1985} were interested in this algorithm
purely as a means to reach the optimal solution, and were
uninterested in statistical considerations which led us to consider
intermediate IRP solutions as regularized isotonic models of
independent interest. \citet{Spou2003} also used \citet
{Maxw1985} to
inspire the partitioning algorithm for the standard isotonic
regression problem, however, they do not make the connection using
\citet{Barl1972}, and also have no statistical interests in mind.

As our analysis and experiments have demonstrated, computation is
not a significant concern with IRP, at least for moderate to large
data sizes. However, overfitting is still a major concern as
dimensionality grows. As demonstrated in Sections
\ref{sdegreesfreedom} and \ref{sperformance}, while IRP offers
partial protection from overfitting through its regularization
behavior, even the first step in the IRP path could already suffer
from high variance in a dimension as low as six. A~key question
pertains to identifying factors affecting this overfitting behavior,
specifically characterizing situations in which the initial IRP
iterations are less prone to overfitting.

\begin{appendix}\label{app}
\section*{Appendix}

\begin{pf*}{Proof of Proposition \ref{propcartvsirp}}
We first rewrite both the IRP partition problem
(\ref{eqoptimalcut}) and the maximal between-group variance
partition problem~(\ref{eqbetweenvarcut}). Assume $V=A\cup B$
and $A\cap B=\{\cdot\}$. Then it is easy to show
$|V|\overline{y}_V=|A|\overline{y}_A+|B|\overline{y}_B$ which gives
$(\overline{y}_A-\overline{y}_V)=-|B|(\overline{y}_B-\overline{y}_V)/|A|$.
The objective function to\vadjust{\goodbreak} (\ref{eqoptimalcut}) can be written
$|B|(\overline{y}_B-\overline{y}_V)-|A|(\overline{y}_A-\overline{y}_V)$
and using the previous relationship can again be rewritten
$2|B|(\overline{y}_B-\overline{y}_V)$. An obvious property of the
optimal IRP cut is that $\overline{y}_B\ge\overline{y}_V$. If we
add this as a~redundant constraint to the IRP partition
(\ref{eqoptimalcut}), then we can find the same optimal partition
by maximizing the square of the objective, that is, maximize
$4|B|^2(\overline{y}_B-\overline{y}_V)^2$ subject to the appropriate
constraints. The objective of the between-group variance partition
(\ref{eqbetweenvarcut}) can be rewritten using the above
relationship as $(|B|+|B|^2/|A|)(\overline{y}_B-\overline{y}_V)^2$.
Then denoting the IRP and maximal between-group variance objectives
by $g^*(A,B)$ and $\tilde{g}(A,B)$, respectively, we have
$g^*(A,B)=4|A||B|\tilde{g}(A,B)/n$ since $|A|+|B|=n$ is constant.
Eliminating the constant $4/n$ gives the first result.

In order to prove the second statement, notice that optimality of (\ref
{eqoptimalcut}) and (\ref{eqbetweenvarcut}) gives $|A^*||B^*|\tilde
{g}(A^*,B^*)\ge|A^*||B^*|\tilde{g}(\tilde{A},\tilde{B})$ and
$\tilde{g}(\tilde{A},\tilde{B})\ge\tilde{g}(A^*,B^*)$ which
implies $|A^*||B^*|\ge|\tilde{A}||\tilde{B}|$. This along with the relation
\[
(|A^*|+|B^*|)^2=|A^*|^2+2|A^*||B^*|+|B^*|^2=|\tilde{A}|^2+2|\tilde
{A}||\tilde{B}|+|\tilde{B}|^2=(|\tilde{A}|+|\tilde{B}|)^2
\]
gives $|A^*|^2+|B^*|^2\leq|\tilde{A}|^2+|\tilde{B}|^2$. We use this
to get the relation
\begin{eqnarray*}
(|A^*|-|B^*|)^2&=&|A^*|^2-2|A^*||B^*|+|B^*|^2\\
&\leq&|\tilde{A}|^2-2|A^*||B^*|+|\tilde{B}|^2\leq|\tilde
{A}|^2-2|\tilde{A}||\tilde{B}|+|\tilde{B}|^2\\
&=&(|\tilde{A}|-|\tilde{B}|)^2,
\end{eqnarray*}
which gives the second result of the proposition.
\end{pf*}
\begin{pf*}{Proof of Theorem \ref{thnoregretcut}}
Divide the blocks in $V$ into three subsets:
\begin{longlist}[(1)]
\item[(1)]$\mathcal{L}$: union of all blocks in $V$ that are ``below'' the
algorithm
cut.
\item[(2)]$\mathcal{U}$: union of all blocks in $V$ that are ``above'' the
algorithm
cut.
\item[(3)]$\mathcal{M}$: union of $K$ blocks in $V$ that get broken by the
cut (note that blocks in $\mathcal{M}$ may be separated by blocks in
$\mathcal{L}$ or
$\mathcal{U}$).
\end{longlist}

Define $M_1$ ($M_K$) to be the minorant (majorant) block in
$\mathcal{M}$. For each~$M_k$ define~$M_{k}^L$ ($M_{k}^U$) as the
groups\vspace*{1pt} in $M_k$ below (above) the algorithm cut. Define
$A_K^L\subseteq\mathcal{L}$ ($A_1^U\subseteq\mathcal{U}$) as the
union\vspace*{1pt} of blocks along the algorithm cut such that
$A_K^L\succ M_K^L$ ($A_1^U\prec M_1^U$). Refer to Figure \ref
{figproof1} for an example of these definitions where
$A_1^U=A_1^L=A_K^U=A_K^L=\{\cdot\}$ for simplicity.

%
\begin{figure}

\includegraphics{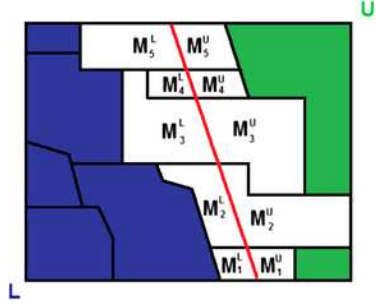}

\caption{Illustration of the proof of Theorem
\protect\ref{thnoregretcut}. Black lines separate blocks. The
diagonal red line through the center demonstrates a cut of Algorithm
\protect\ref{algIRnoregret}. $\mathcal{L}$ is the union of blue
blocks below
the cut and $\mathcal{U}$ is the union of green blocks above the cut.
White blocks are blocks that are potentially split by Algorithm
\protect\ref{algIRnoregret}. These blocks are split into
$M_1^L,\ldots,M_5^L$ below the cut and $M_1^U,\ldots,M_5^U$ above the
cut. In the proof, $M_i=M_i^L\cup M_i^U$ for all$i=1,\ldots,5$. The
proof shows, for example, that if the algorithm splits $M_1$ into
$M_1^L$ and $M_1^U$ according to the defined cut in
(\protect\ref{eqoptimalcutlp}), then there must be no isotonicity violation
when creating blocks from~$M_1^L$ and~$M_1^U$. However, since $M_1$ is
assumed to be a block, there must exist an isotonicity violation
between $M_1^L$ and $M_1^U$, providing a contradiction.}
\label{figproof1}
\end{figure}

We use the above definitions and assumptions to state the following
two consequences that cause a contradiction:
\begin{longlist}[(II)]
\item[(I)]$\overline{y}_{M_1} < \overline{y}_{M_K}$ by optimality (i.e.,
according to KKT conditions) and
isotonicity.
\item[(II)]$\overline{y}_{M_1}>\overline{y}_{V}$ and $\overline
{y}_{M_K}<\overline{y}_{V}$. This is
proven below.
\end{longlist}
(II) implies $\overline{y}_{M_1}>\overline{y}_{M_K}$ which
contradicts (I) and we are left to prove (II). Optimality of blocks
$M_1$ and $M_K$ gives:
\begin{longlist}[(a)]
\item[(a)]$\overline{y}_{M_{1}^L}> \overline{y}_{M_{1}^U}$,
\item[(b)]$\overline{y}_{M_{K}^L} > \overline{y}_{M_{K}^U}$.\vadjust{\goodbreak}
\end{longlist}
The proof for $\overline{y}_{M_1}>\overline{y}_{V}$ is as
follows with two cases:
\begin{longlist}[(2)]
\item[(1)]$A_1^U=\{\cdot\}$: $\overline{y}_{M_1^U}> \overline{y}_{V}$ because
using the algorithm cut in (\ref{eqoptimalcutlp}), we have
\[
\sum_{i\dvtx x_i\in
M_{1}^U}{(y_i-\overline{y}_{V})}>0\quad\Rightarrow\quad
\sum_{i\dvtx x_i\in M_{1}^U}{y_i}>|M_{1}^U|\overline
{y}_{V}\quad\Rightarrow\quad\overline{y}_{M_{1}^U}>\overline{y}_{V}.
\]
The first inequality is true about the cut because there exists no
block below~${M_{1}^U}$ to affect isotonicity. Then using (a), we get
\[
\overline{y}_{M_1^L}>\overline{y}_{M_1^U}>
\overline{y}_{V}\quad\Rightarrow\quad
\overline{y}_{M_1}>\overline{y}_{V}.
\]
\item[(2)]$A_1^U\neq\{\cdot\}$:
$\overline{y}_{M_1}>\overline{y}_{A_1^U}>\overline{y}_{V}$.
The first inequality is due to optimality and the second is again
because the algorithm cut in (\ref{eqoptimalcutlp}) gives
\[
\sum_{i\dvtx x_i\in
A_{1}^U}{(y_i-\overline{y}_{V})}>0\quad\Rightarrow\quad
\sum_{i\dvtx x_i\in A_{1}^U}{y_i}>|A_{1}^U|\overline
{y}_{V}\quad\Rightarrow\quad\overline{y}_{A_{1}^U}>\overline{y}_{V},
\]
which again is possible because no block exists below ${A_{1}^U}$ to
affect isotonicity.
\end{longlist}
The proof for $\overline{y}_{M_K} < \overline{y}_{V}$ is a
similar argument and hence gives (II). The case \mbox{$K=1$} is also
trivially covered by the above arguments. We conclude that the
algorithm cannot cut any block.
\end{pf*}

The following remark is necessary for completeness of the proof of
Theorem \ref{thnoregretcut}.
\begin{remark} \label{remstrictinequalities}
The case of two connected optimal groups having equal means need not
be discussed in Theorem \ref{thnoregretcut}. In this event, the
optimal solution to isotonic regression in not unique. It is
trivial that $M_1$ would not have been split by Algorithm
\ref{algIRnoregret} if
$\overline{y}_{M_1^L}=\overline{y}_{M_1^U}\ne\overline{y}_{V}$.
Otherwise, consider the case
$\overline{y}_{M_1^L}=\overline{y}_{M_1^U}=\overline{y}_{V}$
and assume $M_1$ is a block broken by the cut in
$V$. $M_1^L$ and $M_1^U$ are also possible blocks
whereby $M_1^L\in\mathcal{L}$ and $M_1^U\in\mathcal{U}$, and, hence,
$M_1=M_1^L\cup M_1^U\notin\mathcal{M}$. The same remarks apply to
$M_K$. Thus, the proof still holds if there are multiple isotonic
solutions.
\end{remark}
\begin{remark} \label{remmultipleobservations}
The case of multiple observations at the same coordinates can
be disregarded. To see this, let $J$ be a set of nodes with the same
coordinates. From the constraints, $y_i=y_j$, for all $i,j\in J$ and,
thus, the number of observations can be reduced and all observations in
$J$ fit to the same value $\hat{y}$. Then
\[
\sum_{j\dvtx x_j\in J}{(\hat{y}-y_j)^2}=|J|\mid(\hat
{y}-\overline{y}_J)^2+ \sum_{j\dvtx x_j\in
J}{y_j^2}-\overline{y}_J^2
\]
so that the sum of squared differences over $J$ can be reduced to be a
single weighted squared difference. Problem (\ref{eqIR}) becomes the
weighted isotonic regression problem
%
\begin{equation}\label{eqIRweighted}
\min\Biggl\{ \sum_{i=1}^n{w_i(\hat{y}_i-y_i)^2} \dvtx\hat
{y}_i\leq\hat{y}_j \mbox{ for all } (i,j)\in\mathcal
{I}\Biggr\},
\end{equation}
for which the KKT conditions imply that observations are again divided
into k groups where the fits in each group take the weighted group mean
$\overline{y}^w_V=\sum_{i\dvtx x_i\in V}{(w_iy_i)}/\sum_{i\dvtx x_i\in V}{w_i}$
rather than the group mean. The optimal cut problem (\ref
{eqoptimalcutlp}) changes to have $z_i=w_i(y_i-\overline{y}^w_V)$ and
the above results on IRP generalize easily, noting that now the
weighted algorithm cut implies $\overline{y}^w_{A}> \overline
{y}^w_{V}$ for a group $A$ on the upper side of the cut such that no
group exists below $A$ that could affect isotonicity.
\end{remark}
\begin{pf*}{Proof of Proposition \ref{propcomplexity}}
Any final partition can be represented by a~simple tree. Consider level
$k$ of the tree. Let $p_k\ge0.5$ be the greatest $p$ over levels
$1,\ldots,k-1$ such that a partition of group size $n_k$ into two
groups of size $pn_k$ and $(1-p)n_k$ where $n_k$ is the corresponding
size of the partitioned group. Denote by $L_k$ the largest group
partitioned at iteration $k$ whose size can be bounded by $|L_k|\leq
np_k^k$. We next note that the complexity of solving a problem with $n$
observations is higher than solving 2 problems with $pn$ and $(1-p)n$
observations. Indeed, $n^3=pn^3+(1-p)n^3>p^2n^3+(1-p)^2n^3$. Thus, we
assume that at iteration $k$, we solve only problems of the largest
possible size (rather than several problems of small size). The number
of groups at iteration $k$ can also be bounded by $n/|L_k|$. Denote
by\vadjust{\goodbreak}
$T_{p_k}(k)$ the complexity of partitioning all groups at level $k$. Then
\[
T_{p_k}(k)\leq O\biggl(\frac{n}{|L|}|L|^3\biggr)=O(n|L|^2)\leq
O(n(np_k^k)^2) = O(n^3)p_k^{2k}.
\]
Then denote by $K$ the total number of levels in the partition tree. We have
\[
\sum_{k=1}^K{T_{p_k}(k)} \leq\sum
_{k=1}^K{O(n^3)p_{\max}^{2k}} \leq\sum_{k=1}^\infty
{O(n^3)p_{\max}^{2k}}=O(n^3)\frac{1}{1-p_{\max}^2}.
\]
\upqed\end{pf*}
\end{appendix}

\section*{Acknowledgments}

The authors are grateful to Quentin Stout for drawing our attention to
additional relevant references. We thank the Editor, Associate Editor
and referee for their thoughtful and useful comments. This study makes
use of data generated by the Wellcome Trust Case-Control Consortium. A
full list of the investigators who contributed to the generation of the
data is available from
\href{http://www.wtccc.org.uk}{www.wtccc.org.uk}. We are also very
grateful to David Golan for his help with organizing the WTCCC data for
our experiments.


%

\printaddresses

\end{document}